
\font\ssbig=cmss10 scaled \magstep1  
\font\bfb=cmbx12                     
\font\tfont=cmbxti10
\font\ninerm=cmr9
\baselineskip=24pt


\newfam\vecfam

\textfont\vecfam=\tfont \scriptfont\vecfam=\seveni
\scriptscriptfont\vecfam=\fivei

\def\kms{{\rm\,km\,s^{-1}}}
\def\msun{{\rm\,M_\odot}}

\def\pc{{\rm\,pc}}

\def\yr{{\rm\,yr}}
\def\alv{\alpha_{{\rm vir}}}
\def\mcl{M_{{\rm cl}}}
\def\rcl{R_{{\rm cl}}}
\def\fg{f_{{\rm gas}}}
\def\ecore{\epsilon_{{\rm core}}}

\def\gtorder{\mathrel{\raise.3ex\hbox{$>$}\mkern-14mu
             \lower0.6ex\hbox{$\sim$}}}
\def\ltorder{\mathrel{\raise.3ex\hbox{$<$}\mkern-14mu
             \lower0.6ex\hbox{$\sim$}}}

\newcount\eqnumber
\eqnumber=1
\def\new{{\rm\the\eqnumber}\global\advance\eqnumber by 1}
\def\eqnam#1{\xdef#1{\the\eqnumber}}
%
\newcount\fignumber
\fignumber=1
\def\newfig#1{{\rm\the\fignumber}\xdef#1{\the\fignumber}\global\advance\fignumber by 1}
\def\refto#1{$^{#1}$}           
\def\ref#1{Ref.~#1}                     
\def\Ref#1{#1}                          
\gdef\refis#1{\item{#1.\ }}                     
\def\beginparmode{\endmode
  \begingroup \def\endmode{\par\endgroup}}
\let\endmode=\par
\def\body{\beginparmode}
\def\head#1{                    
  \goodbreak\vskip 0.5truein    
  {\centerline{\bf{#1}}\par}
   \nobreak\vskip 0.25truein\nobreak}
\def\references                 
  {\head{References}            
   \beginparmode
   \frenchspacing \parindent=0pt \leftskip=1truecm
   \parskip=12pt plus 3pt \everypar{\hangindent=\parindent}}
\def\endreferences{\body}

\catcode`@=11
\newcount\r@fcount \r@fcount=0
\newcount\r@fcurr
\immediate\newwrite\reffile
\newif\ifr@ffile\r@ffilefalse
\def\w@rnwrite#1{\ifr@ffile\immediate\write\reffile{#1}\fi\message{#1}}

\def\writer@f#1>>{}
\def\referencefile{
  \r@ffiletrue\immediate\openout\reffile=\jobname.ref%
  \def\writer@f##1>>{\ifr@ffile\immediate\write\reffile%
    {\noexpand\refis{##1} = \csname r@fnum##1\endcsname = %
     \expandafter\expandafter\expandafter\strip@t\expandafter%
     \meaning\csname r@ftext\csname r@fnum##1\endcsname\endcsname}\fi}%
  \def\strip@t##1>>{}}

\def\citeall#1{\xdef#1##1{#1{\noexpand\cite{##1}}}}
\def\cite#1{\each@rg\citer@nge{#1}}	

\def\each@rg#1#2{{\let\thecsname=#1\expandafter\first@rg#2,\end,}}
\def\first@rg#1,{\thecsname{#1}\apply@rg}	
\def\apply@rg#1,{\ifx\end#1\let\next=\relax
\else,\thecsname{#1}\let\next=\apply@rg\fi\next}

\def\citer@nge#1{\citedor@nge#1-\end-}	
\def\citer@ngeat#1\end-{#1}
\def\citedor@nge#1-#2-{\ifx\end#2\r@featspace#1 
  \else\citel@@p{#1}{#2}\citer@ngeat\fi}	
\def\citel@@p#1#2{\ifnum#1>#2{\errmessage{Reference range #1-#2\space is bad.}%
    \errhelp{If you cite a series of references by the notation M-N, then M and
    N must be integers, and N must be greater than or equal to M.}}\else%
 {\count0=#1\count1=#2\advance\count1 by1\relax\expandafter\r@fcite\the\count0,%
  \loop\advance\count0 by1\relax
    \ifnum\count0<\count1,\expandafter\r@fcite\the\count0,%
  \repeat}\fi}

\def\r@featspace#1#2 {\r@fcite#1#2,}	
\def\r@fcite#1,{\ifuncit@d{#1}
    \newr@f{#1}%
    \expandafter\gdef\csname r@ftext\number\r@fcount\endcsname%
                     {\message{Reference #1 to be supplied.}%
                      \writer@f#1>>#1 to be supplied.\par}%
 \fi%
 \csname r@fnum#1\endcsname}
\def\ifuncit@d#1{\expandafter\ifx\csname r@fnum#1\endcsname\relax}%
\def\newr@f#1{\global\advance\r@fcount by1%
    \expandafter\xdef\csname r@fnum#1\endcsname{\number\r@fcount}}

\let\r@fis=\refis			
\def\refis#1#2#3\par{\ifuncit@d{#1}
   \newr@f{#1}%
   \w@rnwrite{Reference #1=\number\r@fcount\space is not cited up to now.}\fi%
  \expandafter\gdef\csname r@ftext\csname r@fnum#1\endcsname\endcsname%
  {\writer@f#1>>#2#3\vskip -0.7\baselineskip\par}}

\def\ignoreuncited{
   \def\refis##1##2##3\par{\ifuncit@d{##1}%
     \else\expandafter\gdef\csname r@ftext\csname r@fnum##1\endcsname\endcsname%
     {\writer@f##1>>##2##3\vskip -0.7\baselineskip\par}\fi}}

\def\r@ferr{\endreferences\errmessage{I was expecting to see
\noexpand\endreferences before now;  I have inserted it here.}}
\let\r@ferences=\references
\def\references{\r@ferences\def\endmode{\r@ferr\par\endgroup}}

\let\endr@ferences=\endreferences
\def\endreferences{\r@fcurr=0
  {\loop\ifnum\r@fcurr<\r@fcount
    \advance\r@fcurr by 1\relax\expandafter\r@fis\expandafter{\number\r@fcurr}%
    \csname r@ftext\number\r@fcurr\endcsname%
  \repeat}\gdef\r@ferr{}\endr@ferences}


\let\r@fend=\endpaper\gdef\endpaper{\ifr@ffile
\immediate\write16{Cross References written on []\jobname.REF.}\fi\r@fend}

\catcode`@=12

\citeall\refto		
\citeall\ref		%
\citeall\Ref		%

\centerline{\bfb Nature, Vol 416, 59 (7 March 2002)}
\medskip
\centerline{\bfb Massive star formation in 100,000 years from turbulent and pressurized molecular clouds}
\medskip
\centerline{\bfb Christopher F. McKee$^{1,2}$ and Jonathan C. Tan$^{2,3}$}
\smallskip    
1. Department of Physics, University of California, Berkeley, CA 94720, USA.

2. Department of Astronomy, University of California, Berkeley, CA 94720, USA.

3. Princeton University Observatory, Peyton Hall, Princeton, NJ 08544, USA.
\bigskip

{\ssbig Massive stars (with mass $m_* \gtorder 8 \msun$) are
fundamental to the evolution of galaxies, because they produce heavy
elements, inject energy into the interstellar medium, and possibly
regulate the star formation rate.  The individual star formation time,
$t_{*f}$, determines the accretion rate of the star; the value of the
former quantity is currently uncertain by many orders of
magnitude\refto{ber96,mcl97,sta00,beh01,oso99,nak00}, leading to other
astrophysical questions. For example, the variation of $t_{*f}$ with
stellar mass dictates whether massive stars can form simultaneously
with low-mass stars in clusters. Here we show that $t_{*f}$ is
determined by conditions in the star's natal cloud, and is typically
$\sim 10^5 \yr$. The corresponding mass accretion rate depends on the
pressure within the cloud---which we relate to the gas surface
density---and on both the instantaneous and final stellar masses.
Characteristic accretion rates are sufficient to overcome radiation
pressure from $\sim 100\msun$ protostars, while simultaneously driving
intense bipolar gas outflows. The weak dependence of $t_{*f}$ on the
final mass of the star allows high- and low-mass star formation to
occur nearly simultaneously in clusters.}
\medskip

Massive stars form in dense molecular {\it clumps} inside molecular
clouds\refto{plu97}. These regions are highly turbulent and are in
approximate virial equilibrium, with comparable values of the
gravitational energy and the kinetic energy. Observed star forming
regions have a virial parameter
$\alv=5\langle\sigma^2\rangle\rcl/G\mcl$ of order unity\refto{bm92},
where $\langle\sigma^2\rangle$ is the mass-averaged one-dimensional
velocity dispersion, $\rcl$ the radius, and $\mcl$ the mass of the
clump. The regions of high-mass star formation studied in ref. 7 are
characterized by masses $\mcl\sim 3800\msun$ and radii $\rcl\sim
0.5\pc$. The corresponding mean column density and visual extinction
are $\Sigma\sim 1\,{\rm g\,cm^{-2}}$ and $A_V=(N_{\rm H}
/2\times10^{21}\,{\rm cm^{-2}})\,{\rm mag}=214\,\Sigma\,{\rm mag}$,
with a dispersion of a factor of a few.  These column densities are
far greater than those associated with regions of low-mass star
formation, but are comparable to the central surface density in the
Orion Nebula Cluster\refto{hil98}, which is about $1\, {\rm
g\,cm^{-2}}$. These regions have very high mean pressures $\overline{P}$:
\eqnam{\eqpbar}
$$ {\overline{P}\over{k_B}}={3\mcl\langle\sigma^2\rangle \over 4\pi\rcl^3k_B}=
   \left({3\pi\fg\alv \over 20 k_B}\right)G\Sigma^2
	\simeq 2.28\times 10^8  \fg \alv \Sigma^2
     ~{\rm K~cm^{-3}},
\eqno(\new)
$$
where $k_B$ is Boltzmann's constant and $\fg$ is the fraction of the
cloud's mass that is in gas, as opposed to stars. They have power-law
density profiles\refto{tak00} $\rho\propto r^{-k_\rho}$, with 
$k_\rho\simeq1.5\pm 0.5$.

Molecular clouds are observed to be inhomogeneous on a wide range of
scales, and numerical simulations show that this is a natural outcome
of supersonic turbulence\refto{vaz00}. Dense, self-gravitating
inhomogeneities that are destined to become stars are termed {\it
cores}.  We assume that the distribution of core masses determines the
resulting distribution of stellar masses (the initial mass function)
and we take this distribution as given.  Because of the disruptive effects
of protostellar outflows, only a fraction $\ecore$ of the total core mass
$M_{\rm core}$ ends up in the star: $m_{*f} = \ecore M_{\rm core}$,
where $m_{*f}$ is the final stellar mass (for binary and other
multiple star systems, $m_{*f}$ is the total mass of stars in the
system).  To estimate $\ecore$ for massive stars, we assume that their
outflows are scaled versions of the outflows from low-mass stars, which is
qualitatively consistent with observation\refto{ric00}.  For low-mass
stars, protostellar outflows typically eject 25-75\% of the core
mass\refto{mat00}. A similar analysis applied to high-mass stars
yields comparable results (J.C.T. \& C.F.M., in prep.), so we
adopt $\ecore=0.5$ as a typical value.

Our fundamental assumption is that star-forming clumps and the cores
embedded within them are each
part of a self-similar, self-gravitating
turbulent structure that is virialized---that is, in approximate
hydrostatic equilibrium---on all scales above the thermal Jeans mass.
The density and pressure are then power laws in radius (for the
pressure, $P\propto r^{-k_P})$, so that the cloud is a polytrope with
$P\propto\rho^{\gamma_p}$. In hydrostatic equilibrium\refto{mcl96},
$k_\rho=2/(2-\gamma_p)$ and $k_P=\gamma_p k_\rho = 2
\gamma_p/(2-\gamma_p)$. Let $c\equiv(P/\rho)^{1/2}\propto
r^{(1-\gamma_p)/(2-\gamma_p)}$ be the effective sound speed; if the
pressure is dominated by turbulent motions, then $c$ is proportional
to the line width. Molecular clouds and cloud cores satisfy a line
width-size relation in which $c$ increases outwards\refto{lar81},
corresponding to $\gamma_p<1$. The equation of hydrostatic equilibrium
gives $M=k_Pc^2r/G$ for the mass inside $r$, and $\rho=Ac^2/(2\pi
Gr^2)$ for the density at $r$, where
$A=\gamma_p(4-3\gamma_p)/(2-\gamma_p)^2$. It is then
immediately possible to determine the properties of a core in terms
of the pressure at its surface 
and the mass of the star that will
form within it (see Fig. 1). 
The radius of a core is then
$r_s=0.074(m_{*f}/30\msun)^{1/2}\Sigma^{-1/2}\pc$;
recall that the typical clump observed in ref. 7
has a radius of 0.5 pc, which is set by the
angular resolution of those observations.  The mean density
in a core is $\bar n_{{\rm
H}}=1.0\times 10^6(m_{*f}/30\msun)^{-1/2}\Sigma^{3/2}\,{\rm
cm^{-3}}$, and the r.m.s. velocity dispersion is
$1.65(m_{*f}/30\msun)^{1/4}
\Sigma^{1/4}~{\rm km\,s^{-1}}$. 
The thermal sound speed at temperature $T$ is $0.3(T/30\, {\rm
K})^{1/2}\,\kms$, and so a high-mass core is dominated by supersonic
turbulence and should be very clumpy.

We now consider the timescale for a star to form in such a core.
On dimensional grounds, we expect the protostellar accretion rate to 
be\refto{sta80}:
\eqnam{\eq:dim}$$ \dot{m}_*=\phi_* {m_* \over t_{\rm ff}},
\eqno(\new)
$$
so long as radiation pressure is not important. Here $m_*$ is the {\it
instantaneous} stellar mass, $t_{\rm ff}=[3\pi/(32G\rho)]^{1/2}$ is
the free-fall time and $\phi_*$ is a dimensionless constant of order
unity. This equation could be violated in the sense that $\phi_*\gg 1$
only in the unlikely case that the star forms from a coherent
spherical implosion; if the star formation is triggered by an external
increase in pressure, $\phi_*$ could be increased somewhat, but
deviations from spherical symmetry in the triggering impulse and in
the protostellar core will generally prevent $\phi_*$ from becoming
too large. It could be violated in the opposite sense that $\phi_*\ll
1$ if the core is magnetically dominated, so that collapse could not
begin until the magnetic field diffused out of the core. However,
magnetic fields are 
not observed to be dominant
in molecular cores\refto{cru99}. Equation (\eq:dim) implies that
the accretion rate, and thus the star formation time $t_{*f}\propto
m_*/\dot{m}_*$, depends weakly on the properties of the ambient
medium, $\dot{m}_*\propto \rho^{1/2}$.

We now show that if the collapse is spherical and self-similar, then
$\phi_*$ is quite close to unity provided that radiation pressure does
not disrupt the flow.  Although we assume the collapse is spherical
far from the star, it will naturally proceed via a disk close to the
star owing to the angular momentum of the accreting material; we
assume this does not limit the flow of matter onto the star as
otherwise the disk would become very massive and gravitationally
unstable.  Shielding by the disk reduces the importance of radiation
pressure on the accretion flow, and allows the formation of massive
stars provided the accretion rate is sufficiently large\refto{jij96}.
Under the assumption of a polytropic structure in hydrostatic
equilibrium, equation (\eq:dim) implies:
\eqnam{\eqmdot}$$ \dot{m}_*=\phi_* \ecore {4 \over \pi \sqrt{3}} 
k_P A^{1/2}{c^3\over G},
\eqno(\new)
$$
where the value of $\rho$ entering $t_{\rm ff}$ is evaluated at the
radius in the initial cloud that just encloses the gas that goes into
the star when its mass is $m_*$.  For the isothermal case
($\gamma_p=1$) and for $\ecore=1$, this reduces to the classic result
of Shu\refto{shu77} with $\phi_*=0.975\pi
\sqrt{3}/8=0.663$. McLaughlin \& Pudritz\refto{mcl97} show that the
accretion rate $\dot m_*$ is proportional to
$t^{3-3\gamma_p}$, so that for $\gamma_p<1$, as observed, the
accretion rate {\it accelerates}\refto{mcl97,beh01}. As discussed in
ref. 2, termination of the accretion breaks the self-similarity once
the expansion wave encloses $m_{*f}$, which occurs at $t\sim
0.45t_{*f}$. Thereafter, the relation $\dot{m}_*\propto
t^{3-3\gamma_p}$ becomes approximate, but the approximation should be reasonable good\refto{mcl97}. The star-formation time is then given by:
\eqnam{\eq:tsf}
$$ t_{*f} = {(4-3\gamma_p)\over \phi_*}t_{\rm ff}.
\eqno(\new)
$$
Evaluating $\phi_*$ (see Fig. 1), we conclude that
$\dot{m}_*\simeq m_*/t_{\rm ff}$ to within a factor 1.5 for spherical
cores in which the effective sound speed increases outward.

We can express the accretion rate in terms of the mean pressure of the
clump in which the star forms, $\bar P$ (eq. 1), and the final mass of
the star, $m_{*f}$,
\eqnam{\eqmdotSigma}
$$ \dot{m}_* = 4.75 \times 10^{-4} \ecore^{1/4}
(\fg\phi_P\alv)^{3/8} \left({m_{*f} \over 30\msun}\right)^{3/4} \Sigma^{3/4} 
\left({m_*\over m_{*f}}\right)^j\,{\rm M_\odot\,yr^{-1}},
\eqno(\new)
$$
where $\phi_P\equiv P_s/ \overline{P}$ is the ratio of the core's
surface pressure to the mean pressure in the clump and $j\equiv
3(2-k_\rho)/[2(3-k_\rho)]$; for $k_\rho=3/2$, $j=1/2$.
This typical accretion rate is large because the core is embedded in a
high-pressure environment, so that in hydrostatic equilibrium
it has a high density and short free-fall time.
Because massive cores are turbulent and clumpy, we expect the accretion
rate to exhibit large fluctuations. Whereas clumpiness is an attribute
of massive star-forming regions in our model, it is not a
prerequisite, as suggested in ref. 3.
The star formation time is:
\eqnam{\eq:tsfSigma}
$$ t_{*f} = 1.26 \times 10^{5} 
\ecore^{-1/4} 
(\fg\phi_P\alv)^{-3/8}
\left({m_{*f}\over30\msun}\right)^{1/4} 
\Sigma^{-3/4}\,{\rm yr}.
\eqno(\new)
$$
The weak dependence of the star formation time on the final stellar
mass means that low-mass and high-mass stars can form approximately at
the same time.  Furthermore, $t_{*f}$ is small compared to the
estimated timescale $\sim 10^6\,{\rm yr}$ for cluster
formation\refto{pal99}.  With the fiducial values of the parameters, a
$100\msun$ star forming in a clump with $\Sigma=1\, {\rm g\,cm^{-2}}$
has a final accretion rate of $1.1\times 10^{-3}\,{\rm
M_\odot\,yr^{-1}}$ and a star formation time of $1.8\times 10^5\yr$.

	A necessary condition for massive star formation is that the
ram pressure associated with accretion exceed the radiation pressure
at the point where the dust in the infalling gas is
destroyed\refto{wol87}.  For a 100 $M_\odot$ star, this
requires\refto{jij96} $\dot m_*\gtorder 6\times 10^{-4}\;M_\odot$
yr$^{-1}$, which is satisfied for $\Sigma\gtorder 0.5$ g cm$^{-2}$.
Once the core has formed a star, the star can continue to grow by
Bondi-Hoyle accretion\refto{bon01} provided $m_*\ltorder 10\;\msun$ so
that radiation pressure does not prevent the focussing of gas
streamlines in the wake of the star. The Bondi-Hoyle accretion rate
$\dot{m}_*= 2.2\times10^{-7} (M_{\rm
cl}/4\times10^3\msun)^{-5/4}\Sigma^{3/4} (m_*/10\msun)^2
\msun\yr^{-1}$ (CFM \& JCT, in prep.) is so low 
that this process does not significantly alter the stellar mass under
the conditions observed in the clumps of ref.~7.

Direct comparison of our results with observation is difficult because
the actual masses of massive protostars are poorly determined.  Our
approach is to predict the properties of some well-studied
massive protostars in terms of their bolometric luminosities. The
bolometric luminosity $L_{\rm bol}$ has contributions from
main-sequence nuclear burning $L_{\rm ms}$, deuterium burning $L_D$,
and accretion $L_{\rm acc}$.  The accretion luminosity $L_{\rm
acc}=f_{\rm acc} G m_*\dot{m}_*/r_*$, where $f_{\rm acc}$ is a factor
of order unity accounting for energy radiated by an accretion disk,
advected into the star or converted into kinetic energy of outflows,
and where the stellar radius $r_*$ may depend sensitively on the
accretion rate $\dot{m}_*$. Massive stars join the main sequence
during their accretion phase at a mass that also depends on the
accretion rate\refto{pal92}.  To treat accelerating accretion rates,
we have developed a simple model for protostellar evolution based on
that of Nakano et al.\refto{nak00,nak95} The model accounts for the
total energy of the protostar as it accretes and dissociates matter
and, if the central temperature $T_c\gtorder 10^6\,{\rm K}$, burns
deuterium.  We have modified this model to include
additional processes, such as deuterium shell burning, and we have
calibrated these modifications against the more detailed calculations
of Palla \& Stahler\refto{pal91,pal92}.

Our model allows us to make predictions for the masses and accretion
rates of embedded protostars that are thought to power hot molecular
cores (CFM \& JCT, in prep.). Figure 2 compares our
theoretical tracks with the observed bolometric luminosities of
several sources.  We find that uncertainties in the value of the
pressure create only small uncertainties in $m_*$ for $L_{\rm
bol}\gtorder {\rm few} \times 10^4\,{\rm L_\odot}$.

The infrared and submillimeter spectra of accreting protostars and
their surrounding envelopes have been modelled in ref. 5., modelling
the same sources shown in Fig. 2. We note that uncertainties in the
structure of the gas envelope and the possible contributions from
additional surrounding gas cores or diffuse gas will affect the
observed spectrum. Comparing results, our inferred stellar masses are
similar, but our accretion rates are systematically smaller by factors
of $\sim 2-5$. The modelled\refto{oso99} high accretion rates of $\sim
10^{-3}\,{\rm M_\odot\,yr^{-1}}$ for stars with $m_*\sim 10\msun$
would be difficult to achieve unless the pressure was increased
substantially; for example, if the stars are destined to reach
$m_{*f}\sim 30\msun$, pressure increases of a factor $\sim 40$ 
are required.

\noindent
{\ninerm Received 17 September 2001; accepted 2 January 2002}

\references
{
\refis{wol87} Wolfire, M. G. \& Cassinelli, J. Conditions for the formation of 
massive stars. {\it Astrophys.\ J.} {\bf 319}, 850-867 (1987).
\par

\refis{ber96} Bernasconi, P. A. \& Maeder, A. About the absence of a proper 
zero age main sequence for massive stars. {\it Astron. Astrophys.} {\bf 307}, 829-839 (1996).\par

\refis{mcl97} McLaughlin, D. E. \& Pudritz, R. E. Gravitational Collapse and 
Star Formation in Logotropic and Nonisothermal Spheres. {\it Astrophys. J.} {\bf 476}, 
750-765 (1997).\par

\refis{sta00} Stahler, S. W., Palla, F. \& Ho, P. T. P. The Formation of 
Massive Stars. Protostars \& Planets IV, eds. Mannings, V., Boss,
A. P. \& Russell, S. S. 
(University of Arizona Press, Tucson), 327-351 (2000).\par

\refis{vaz00} V\'azquez-Semadeni, E., Ostriker, E. C., Passot, T., Gammie, C. F. \& Stone, J. M. Compressible MHD Turbulence: Implications for Molecular Cloud and Star Formation. Protostars \& Planets IV, eds. Mannings, V., Boss, A. P. \& Russell, S. S. 
(University of Arizona Press, Tucson), 3-28 (2000).\par

\refis{ric00} Richer, J. S., Shepherd, D. S., Cabrit, S., Bachiller, R.,
\& Churchwell, E. Molecular Outflows from Young Stellar Objects.
Protostars \& Planets IV, eds. Mannings, V., Boss,
A. P. \& Russell, S. S. 
(University of Arizona Press, Tucson), 867-894 (2000).\par

\refis{beh01} Behrend, R. \& Maeder, A. Formation of massive stars by growing 
accretion rate. {\it Astron. Astrophys.} {\bf 373}, 190-198 (2001).\par

\refis{oso99} Osorio, M., Lizano, S. \& D'Alessio, P. Hot Molecular Cores and 
the Formation of Massive Stars. {\it Astrophys. J.} {\bf 525}, 808-820 (1999).\par

\refis{nak00} Nakano, T., Hasegawa, T., Morino, J.-I. \& Yamashita, T. 
Evolution of Protostars Accreting Mass at Very High Rates: Is Orion IRc2 a Huge 
Protostar? {\it Astrophys. J.} {\bf 534}, 976-983 (2000).\par

\refis{sta80} Stahler, S. W., Shu, F. H. \& Taam, R. E. The evolution of 
protostars. I - Global formulation and results. {\it Astrophys. J.} {\bf 241}, 637-654 (1980).\par

\refis{cru99} Crutcher, R. M. Magnetic Fields in Molecular Clouds: 
Observations Confront Theory. {\it Astrophys. J.} {\bf 520}, 706-713 (1999).\par


\refis{jij96} Jijina, J. \& Adams, F. C. Infall Collapse Solutions in the 
Inner Limit: Radiation Pressure and Its Effects on Star Formation. {\it Astrophys. J.} {\bf 462}, 
874-887 (1996).\par


\refis{tak00} van der Tak, F. F. S., van Dishoeck, E. F., Evans, N. J. (II) \& 
Blake, G. A. Structure and Evolution of the Envelopes of Deeply Embedded Massive Young 
Stars. {\it Astrophys. J.} {\bf 537}, 283-303 (2000).\par

\refis{mcl96} McLaughlin, D. E. \& Pudritz, R. E. A Model for the Internal 
Structure of Molecular Cloud Cores. {\it Astrophys. J.} {\bf 469}, 194-208 (1996).\par

\refis{lar81} Larson, R. B. Turbulence and star formation in molecular 
clouds. {\it Mon. Not. R. Astron. Soc.} {\bf 194}, 809-826 (1981).\par

\refis{mat00} Matzner, C. D. \& McKee, C. F. Efficiencies of Low-Mass Star and 
Star Cluster Formation. {\it Astrophys. J.} {\bf 545}, 364-378 (2000).\par

\refis{shu77} Shu, F. H. Self-similar collapse of isothermal spheres and star 
formation. {\it Astrophys. J.} {\bf 214}, 488-497 (1977).\par

\refis{bm92} Bertoldi, F. \& McKee, C. F. Pressure-confined clumps in 
magnetized molecular clouds. {\it Astrophys. J.} {\bf 395}, 140-157 (1992).\par

\refis{plu97} Plume, R., Jaffe, D. T., Evans, N. J. (II), Martin-Pintado, J. 
\& Gomez-Gonzalez, J. Dense Gas and Star Formation: Characteristics of Cloud Cores 
Associated with Water Masers. {it Astrophys. J.} {\bf 476}, 730-749 (1997).\par


\refis{hil98} Hillenbrand, L. A. \& Hartmann, L. W. A Preliminary Study of the 
Orion Nebula Cluster Structure and Dynamics. {\it Astrophys. J.} {\bf 492}, 540-553 (1998).\par


\refis{pal92} Palla, F. \& Stahler, S. W. The evolution of intermediate-mass 
protostars. II - Influence of the accretion flow. {\it Astrophys. J.} {\bf 392}, 667-677 (1992).\par

\refis{nak95} Nakano, T., Hasegawa, T. \& Norman, C. The Mass of a Star Formed 
in a Cloud Core: Theory and Its Application to the Orion A Cloud. {\it Astrophys. J.}
{\bf 450}, 183-195 (1995).\par

\refis{pal91} Palla, F. \& Stahler, S. W. The evolution of intermediate-mass 
protostars. I - Basic results. {\it Astrophys. J.} {\bf 375}, 288-299 (1991).\par

\refis{pal99} Palla, F. \& Stahler, S. W. Star Formation in the Orion Nebula Cluster.
{\it Astrophys. J.} {\bf 525}, 772-783 (1999).\par

\refis{sch92} Schaller, G., Schaerer, D., Meynet, G. \& Maeder, A. New grids 
of stellar models from 0.8 to 120 solar masses at Z = 0.020 and Z = 0.001. 
{\it Astron. Astrophys. Supp.} {\bf 96}, 269-331 (1992).\par

\refis{hun98} Hunter, T. R. et al. G34.24+0.13MM: A Deeply Embedded 
Proto-B-Star. {\it Astrophys. J.} {\bf 493}, L97-L100 (1998).\par
\refis{mol98} Molinari, S., Testi, L., Brand, J., Cesaroni, R. \& Palla, F. 
IRAS 23385+6053: A Prototype Massive Class 0 Object. {\it Astrophys. J.} {\bf 505}, L39-L42 
(1998).\par

\refis{kau98} Kaufman, M. J., Hollenbach, D. J. \& Tielens, A. G. G. M. High-
Temperature Molecular Cores near Massive Stars and Application to the Orion 
Hot Core. {\it Astrophys. J.} {\bf 497}, 276-287 (1998).\par

\refis{wyr99} Wyrowski, F., Schilke, P., Walmsley, C. M. \& Menten, K. M. Hot 
Gas and Dust in a Protostellar Cluster near W3(OH). {\it Astrophys. J.} {\bf 514}, L43-L46 
(1999).\par


\refis{bon01} Bonnell, I. A., Bate, M. R., Clarke, C. J. \& Pringle, J. E. Competitive accretion in embedded stellar clusters. {\it Mon. Not. R. Astron. Soc.} {\bf 323}, 785-794 (2001).\par

}
\endreferences
\endmode
\bigskip
\bigskip
\bigskip
\bigskip
\bigskip
\par\noindent ACKNOWLEDGEMENTS. We thank Steve Stahler, Ralph Pudritz, 
Malcolm Walmsley and Mark Krumholz for discussions. This work was
supported by the NSF, by NASA (which supports the Center for Star
Formation Studies) and (for J.C.T.) by a Spitzer-Cotsen fellowship from
Princeton University.

\bigskip
\bigskip
\medskip
Correspondence should be addressed to 
CFM (e-mail: cmckee@mckee.berkeley.edu) or to JCT (e-mail:
jt@astro.princeton.edu).

\bigskip
\bigskip
\bigskip
\bigskip
\bigskip
\bigskip
\bigskip
\bigskip
\bigskip

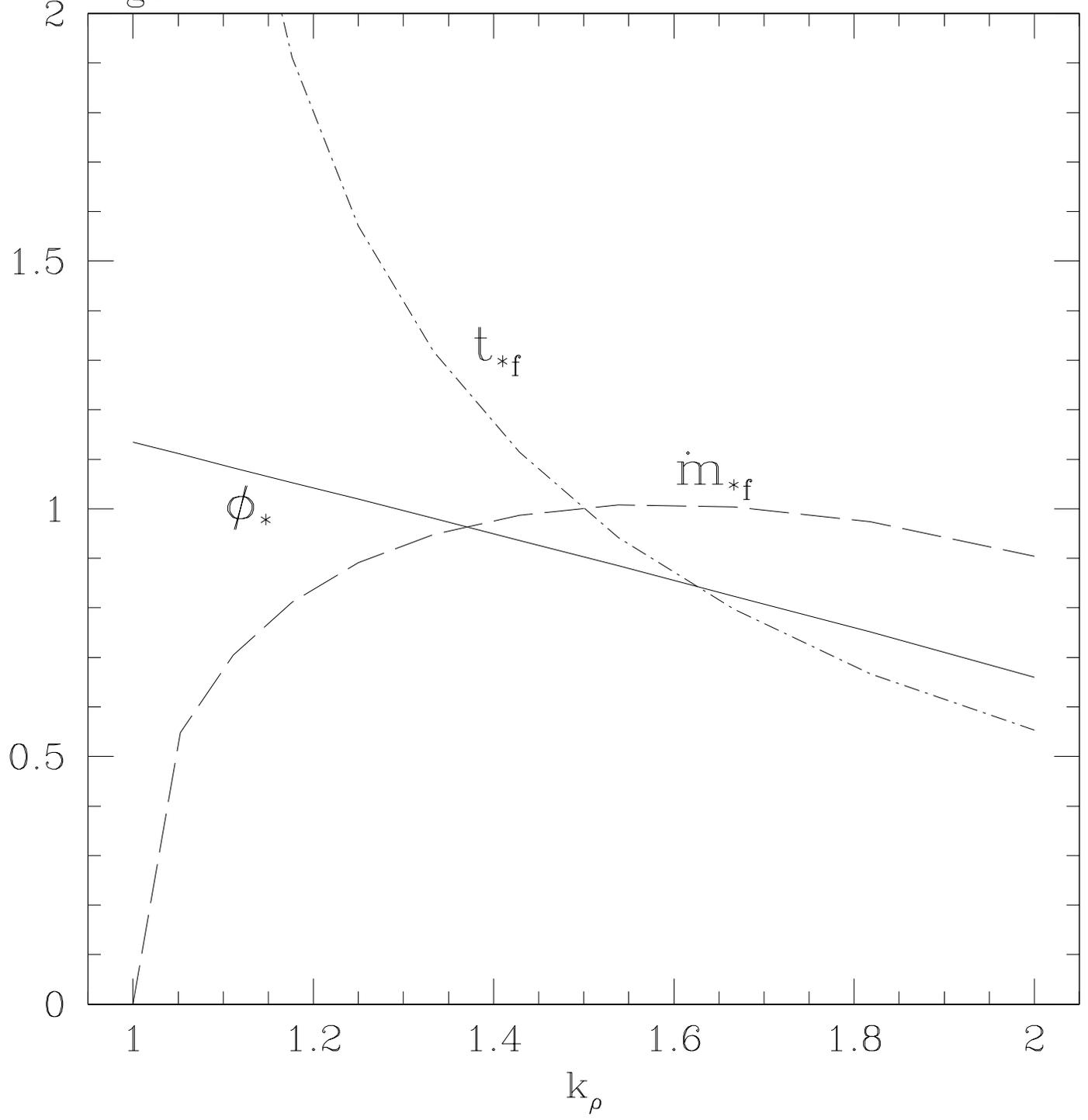
{\bf Fig. 1. Variation of model parameters and results with $k_\rho$.}  Let
$P_s=\phi_P\overline{P}$ be the surface pressure of a core.  The
properties at the surface of the core are given by
$r_s=[AGm_{*f}^2/(2\pi k_P^{2}\ecore^{2}P_s)]^{1/4}$, $\rho_{s} =
[Ak_P^2\ecore^2P_s^3/(2\pi G^3 m_{*f}^2)]^{1/4}$, and $c_s=[2\pi G^3
m_{*f}^2 P_s/(A k_P^2\ecore^2)]^{1/8}$.  We anticipate that the
overall star formation efficiency in the clump will be relatively
high, so in equation~(\eqpbar) we adopt $\fg=2/3$ as a fiducial value.
We estimate $\phi_P= P_s/\overline{P} \simeq 2$ (C.F.M.
\& J.C.T., in prep.), and we
set $\alv$ equal to unity\refto{plu97}.  We take $k_\rho=1.5$ as a
typical value, corresponding to $\gamma_p=2/3$, $k_P=1$ and $A=3/4$.
Following equation (\eqmdot), we evaluate $\phi_*$ using the results
of McLaughlin \& Pudritz\refto{mcl97}, $\phi_*=\pi\sqrt{3}
[(2-\gamma_p)^2(4-3\gamma_p)^{(7-6\gamma_p)/2}
8^{(3\gamma_p-5)/2}m_0]^{1/(4-3\gamma_p)}$, where $m_0$ is tabulated
in their paper.  Over the entire range of $\gamma_p$ and $k_\rho$
relevant to molecular clouds ($0\leq\gamma_p\leq1$, $1\leq k_\rho\leq
2$), $\phi_*\simeq 1.62-0.48 k_\rho$ to within about 1\%.  The star
formation time decreases from $3.5t_{\rm ff}$ to $1.5t_{\rm ff}$ as
$\gamma_p$ varies from 0 to 1.  The variation of $t_{*f}$ and
$\dot{m}_{*f}$ relative to the $k_\rho=1.5$ case is also shown.  Note
that the singular polytropic model in hydrostatic equilibrium breaks
down for $k_\rho =1$, $\gamma_p=0$, since then the pressure gradient
vanishes ($k_P=0$).

\medskip

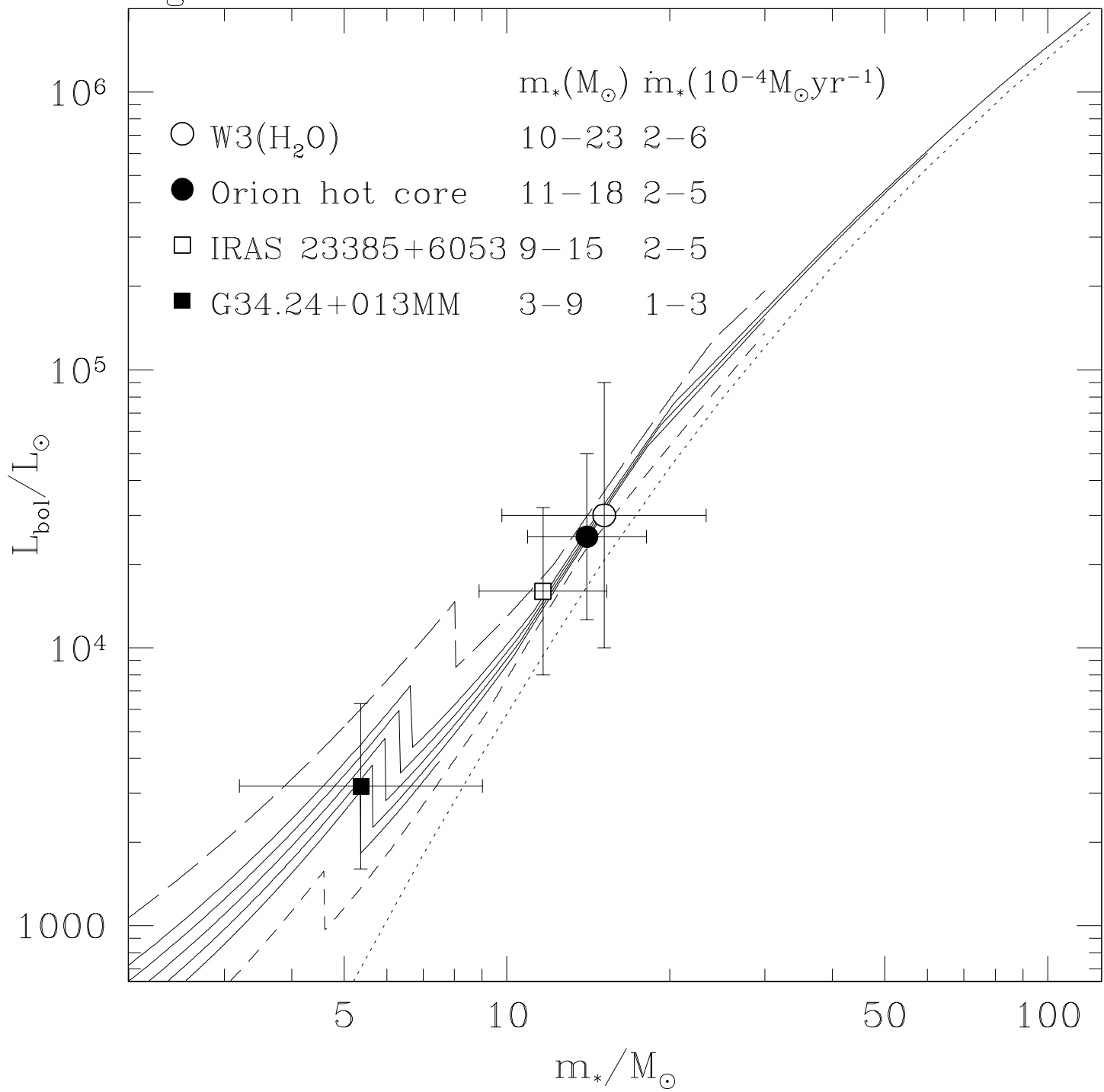
{\bf Fig. 2. Derived properties of nearby massive protostars.} {\it
Solid} lines show the predicted evolution in luminosity of protostars
(including their accretion disks, but allowing for the powering of
protostellar outflows so that $f_{\rm acc}=0.5$; J.C.T. \& C.F.M., in
prep.)  of final mass 7.5, 15, 30, 60 and 120~${\rm M_\odot}$
accreting from cores with $k_\rho=1.5$ embedded in a $\Sigma=1\,{\rm
g\,cm^{-2}}$ clump, typical of Galactic regions\refto{plu97}. The
luminosity step occurring at around 5 to 7~${\rm M_\odot}$, depending
on the model, corresponds to the onset of deuterium shell burning,
which swells the protostellar radius by a factor of about two and thus
reduces the accretion luminosity by the same factor.  The {\it dashed}
and {\it long dashed} lines show a 30~${\rm M_\odot}$ star forming in
a clump with mean pressure ten times smaller and larger than the
fiducial value, respectively.  The {\it dotted} line shows the zero
age main sequence luminosity from ref. 26. Four observed hot
molecular cores are shown: G34.24+0.13MM\refto{hun98}, IRAS
23385+6053\refto{mol98}, Orion Hot Core\refto{kau98} and W3(${\rm
H_2O}$)---the Turner-Welch object\refto{wyr99}. The vertical error bar
illustrates the uncertainty in the bolometric luminosity. The
horizontal error bar shows the corresponding range of allowed values
of $m_*$ for the $\Sigma=1\,{\rm g\,cm^{-2}}$ models. These values and
the constraints on $\dot{m}_*$ are listed here for each source.

\vfill\eject

\bye